\documentclass[12pt]{iopart}
\usepackage{latexsym}
\usepackage{amsfonts}
\usepackage{epsfig}
\usepackage{setspace} 
\usepackage{graphicx}
\begin{document}

\title{Dynamically Slow Processes in Supercooled Water Confined Between Hydrophobic Plates}

\author{Giancarlo Franzese} 

\address{Departamento de F\'isica Fundamental,
Universidad de Barcelona, Diagonal 647, Barcelona 08028, Spain}
\ead{gfranzese@ub.edu}

\author{Francisco de los Santos}
\address{Departamento de Electromagnetismo y F{\'\i}sica de la
Materia, Universidad de Granada, Fuentenueva s/n, 18071 Granada,
Spain}
\ead{fdlsant@ugr.es}

\begin{abstract}
We study the dynamics of water confined between
hydrophobic flat surfaces at low temperature. At different pressures, we
observe different behaviors that we understand in terms of
the hydrogen bonds dynamics. At high pressure, the formation of the
open structure of the hydrogen bond network is inhibited and the
surfaces can be rapidly dehydrated by decreasing the temperature. At
lower pressure the rapid ordering of the  hydrogen bonds generates
heterogeneities that are responsible for strong non-exponential
behavior of the correlation function, but with no strong increase of
the correlation time. At very low pressures, the gradual formation of
the hydrogen bond network is responsible for the large increase of the
correlation time and, eventually, the dynamical arrest of the system
and of the dehydration process.
\end{abstract}

\maketitle

\section{Introduction}

\subsection{Supercooled Water and Polyamorphism}

Supercooled water is a metastable state of the liquid water with
respect to the crystal ice and occurs when H$_2$O is cooled slowly
below the freezing point. It can be easily prepared by cooling
relatively pure commercial water and is commonly observed
in nature, especially under confinement as for example in plants where
water is still liquid at $-47^o$C. It may
crystallize into ice by catalysts, such as an ice crystal, or by a
large exchange of mechanical energy, such as a strong
vibration. However, without catalysts or external work, it is a
useful example of a liquid whose dynamics slows down when the
temperature decreases toward the homogeneous nucleation temperature
$T_H$ (with $T_H=-41^o$C at 1 atm and $T_H=-92^o$C at 2000 atm)
\cite{debenedetti_stanley}. The study  of supercooled water is
important in the long-lasting efforts to understand the physics of the
hydrogen bonds, which characterize water and are the origin of its
unique properties \cite{KumarFS2008}.

Below $T_H$ water crystallizes, but a fast quench  from ambient $T$ to
$T<T_X=-123^o$C (at 1 atm) freezes the liquid into a disordered (not
crystalline) ice, leading to the formation of (amorphous) glassy water
\cite{Loerting_Giovambattista}. By increasing the pressure it has been
observed that glassy water changes from low density amorphous
(LDA) to high density amorphous (HDA) \cite{MISHIMA85} and from HDA to
very-high density amorphous (VHDA) \cite{Loerting01}. Therefore, water
shows polyamorphism and this phenomenon has been related to the
possible existence of more than one liquid phase of water
\cite{llcp}. Other thermodynamically consistent scenarios have been
proposed, such as a first-order liquid-liquid phase transition with no
critical point  \cite{Angell2008} or the absence of liquid-liquid
phase transition itself  \cite{Sastry1996}, and their experimental
implications are currently debated \cite{FranzeseSCKMCS2008}. 

\subsection{Confined Water}

Confinement is an effective way to suppress the crystal
nucleation. Water molecules in plants, rocks, carbon nanotubes,
living cells or on the surface of a protein, are confined to the
nanometer scale. Many recent experiments show a rich phenomenology
which is the object of an intense debate (see for example
\cite{FLMZC2009,Ricci2009,Soper2008,Bellissent2008,Angell2008,GLRD2008,Vogel2008,
  PKS2008,GR2007,MBBCMC2007,CLFBFM2006}). The relation between the
properties of confined and bulk supercooled water is not fully
understood, therefore it is useful to elaborate handleable models that 
could help in developing a consistent theory. 

\subsection{Models for Water}

Since
the experiments in the supercooled region are difficult to perform,
numerical simulations are playing an important role in the
interpretation of the data. These simulations are based on models with
different degree of detailed description and complexity. Among the
most commonly used, there are the ST2 \cite{ST2} (two positive and two
negative charges in tetrahedral position plus a central Lennard-Jones
or LJ), the SPC and its modification SPC/E (three negative and one
positive charges plus a central LJ) \cite{SPC/E}, the TIP3P, TIP4P,
and TIP5P with three, four and five charges, respectively \cite{TIP},
the models with internal degrees of freedom \cite{Ferguson95},
polarizability and three body interactions \cite{3body}, and 
{\it ab initio} models for the evaluation of the significance of
quantum effects \cite{MQ}.  

Models with increasing details lead to improved accuracy in the
simulations, but suffer from important technical limitations in time
and size of the computation. Due to their complexity, an analysis of
how each of their internal parameter affects the properties of the
model is in general very computationally expensive and a theoretical
prediction of their behavior is unfeasible, limiting their capacity to   
offer an insight into unifying concepts in terms of universal
principles and hampering their ability to study thermodynamics
properties.  
 
On the other hand, simplified theoretical models, such as the Ising
model, can be very drastic in their approximations, but are able to
offer a general view on  global, and possibly universal, issues such
as thermodynamics and dynamics. They can be studied analytically or by
simulations, providing theoretical predictions that can be tested by
numerical computation. Therefore, they offer a complementary approach
to the direct simulation of detailed models. Their characterization in
terms of few internal parameters make them more flexible and
versatile, offering unifying views of different cases. Their study by
theoretical methods, such as mean field, leads to analytic functional
relations for their properties, allowing to predict their behavior
over all the possible values of the external parameters. Their reduced
computational cost permits direct test of the theoretical forecasts in
any condition. Furthermore, their study helps in defining improved
detailed models for greater computational efficiency. 
Therefore, there is a
strong need for simplified models of water that can keep the physical
information and provide a more efficient simulation tool.

In the following we will describe a simple model for water in two
dimensions (2D). The model can be easily extended to 3D, however we
consider here only the 2D case. It can be thought as a monolayer of
water between two flat hydrophobic plates where the water-wall
interaction is purely repulsive (steric hard-core exclusion).  It has
been shown that  the properties of confined water are only weakly
dependent on the details of the confining potential between smooth
walls and that the plate separation is the relevant parameter
determining whether or not a crystal forms \cite{KSBS2007}. The case
we consider here corresponds to the physical situation in which the
distance between the plates inhibits the crystallization. 

\section{Cooperative Cell Model Confined Between Partially Hydrated
  Hydrophobic  Plates} 

\subsection{General Definition of the Model}

Cell models with variable volume are particularly suitable to describe
fluids at constant pressure $P$, temperature $T$ and number of
molecules $N$ as in the majority of experiments with water. Here we
consider a cell model based on the idea of including a three-body
interaction, as well as an isotropic and a directional term for the
hydrogen bonds \cite{FS2002,FS2007}. The three-body term takes into
account the cooperative nature of the hydrogen bond. This model
reproduces the experimental phase diagram of liquid water
\cite{FMS2003,KFS2008} and, by varying its parameters, the different
scenarios for the anomalies of water  \cite{MSSSF2009,Stokely08}.  

We consider water confined between two hydrophobic plates within a
volume $V$  
that accommodates $N$ molecules. To minimize the boundary effects, we
consider  double periodic conditions. We assume that the distance
between the plates is such that water forms a monolayer on a 
square surface, similar to the one observed in molecular dynamics
simulations for TIP5P water, where each water molecule  has four
nearest neighbors  \cite{KBSGS2005}. 
The system is divided into $L^2$ cells of equal volume $v=V/L^2$.
Differently from previous works on this model \cite{FYS2000, FS2002,
  FS-PhysA2002,
  FMS2003,FS2007,FranzeseSCKMCS2008,KumarFS2008,KFS2008,MSSSF2009,Stokely08},  
here we consider partially hydrated walls with a hydration level
corresponding to a water concentration of 75\% with respect to the
available surface of one wall. This concentration is well above the
site percolation threshold (59.3\%) on a square lattice
\cite{site-bond}.  
We assume that in each cell $i=1, 2, \ldots L^2$
there is at most one water molecule and we assign the variable $n_i=1$
if the cell is occupied, and $n_i=0$ otherwise.  

The water-water interaction is described by three terms. One takes
into account all the isotropic contributions (short-range repulsion of
the electron clouds, weak attractive van der Waals or London
dispersion interaction, strong attractive isotropic component of the
hydrogen bond interaction). This term is given by  
 a hard-core repulsive at distance volume $R_0=v_0^{1/2}$ plus a LJ
 potential  
\begin{equation}
U(r)=  \left\{
\begin{array}{lll}
\infty  & {\rm for}  &\ r\leq R_0 \\
\displaystyle \epsilon \left[ \left( \frac{R_0}{r}\right)^{12}-\left(
  \frac{R_0}{r}\right)^6\right] 
& {\rm for} & \ r>R_0,
\end{array}\right.
\end{equation}
where $r$ is the distance between two molecules. 

The second term takes into account the directionality of the hydrogen
bonds interaction. Each molecule participates in at most four hydrogen
bonds when each donor O-H forms an angle with the acceptor O that
deviates from linear (180$^o$) within 30$^o$ or 40$^o$
approximately. To incorporate this feature to the model, we associate
to each water molecule $i$ four indices $\sigma_{ij} \in [1,2, \ldots
  q]$, one for each possible hydrogen bond with the nearest neighbors
molecules $j$. The parameter $q$ is chosen by selecting 30$^o$ as
maximum deviation from the linear bond, hence $q=180/30\equiv 6$. With
this choice, every molecule has $q^4=6^4\equiv 1296$ possible
orientations. 
This directional interaction can be accounted for by the energy term
\begin{equation}
 E_J=-J \sum_{\langle i, j  \rangle} n_i n_j
 \delta_{\sigma_{ij},\sigma_{ji}}=-J N_{HB}, 
\end{equation}
where the sum runs over all nearest-neighbor cells, $J>0$ is the
energy associated to the formation of a hydrogen bond,  and $N_{HB}$
is the total number of hydrogen bonds in the system.  

As a consequence of their directionality, the hydrogen bonds induce a
rearrangement of  the molecules in an open network (tetrahedral in
3D), in contrast with the more dense arrangement that can be observed
at higher $T$ or higher $P$ \cite{BBRPSA2007}. This effect can be
included in the model by associating a small volume contribution
$v_{HB}$, due to the formation of each bond, to the total volume of the system
\begin{equation}
 V=V_0+N_{HB}v_{HB}
 \end{equation}
 where $V_0=N v_0$ is the volume of the liquid with no hydrogen bonds. 

Finally, experiments show that the relative orientations of the
hydrogen bonds of a water 
molecule are correlated \cite{intramol}. As an effect of the
cooperativity between the hydrogen bonds, the O--O--O angle
distribution becomes sharper at lower $T$ \cite{Ricci2009}. Therefore,
the accessible $q$ values are reduced. We include this effect in the
model by the energy term 
\begin{equation}
 E_{\sigma}=-J_{\sigma}\sum_i n_i \sum_{(k,l)_i} \delta_{\sigma_{ik},\sigma_{il}},
\end{equation}
where $J_{\sigma}$ is the energy gain due to the cooperative ordering
of the hydrogen bonds, 
$(k,l)_i$ stands for the 6 different pairs of bond indices (arms) of a
molecule $i$. At low $T$ this terms induces the reduction of
accessible values of $q$ (orientational ordering).  
In the limiting case $J_{\sigma}=0$ there is no correlation between
bonds \cite{Sastry1996}. This case corresponds to consider as
negligible the observed change in O--O--O angle distribution with
$T$. 
The case $J_\sigma \gg 1$ corresponds to a drastic reduction of orientational states for the molecule from $q^4$ to $q$, with all the relative orientation between the hydrogen bonds kept fixed independently from $T$ and $P$.

\subsection{The Parameters Dependence of Phase Diagram of the Model}

The complete enthalpy for the fluid reads
\begin{equation}
H=U(r)+E_J+E_{\sigma}+PV.
\label{enthalpy}
\end{equation}
Each of the energy terms of (\ref{enthalpy}) provides an
energy scale, and depending on their relative importance
different physical situations emerge. For instance, the choice 
$J < \epsilon$ guarantees that the hydrogen bonds are mainly formed in the liquid phase. 
As showed in \cite{Stokely08}, when $J_\sigma=0$ the model gives rise
to the scenario with no liquid-liquid coexistence \cite{Sastry1996};
at larger $J_\sigma/J$ the model displays the scenario with a  
liquid-liquid critical point at positive pressure \cite{llcp}; further
increase of $J_\sigma/J$ brings to a liquid-liquid critical point at
negative pressure \cite{Tanaka96}; finally more increase of
$J_\sigma/J$  
brings to the scenario where the liquid-liquid phase transition
extends to the negative pressure of the limit of stability of the
liquid phase \cite{Angell2008}. 

The term
\begin{equation}
E_J+PV\equiv -(J-Pv_{HB})N_{HB}+PV_0
\label{p_term}
\end{equation}
is proportional to the number of hydrogen bonds $N_{HB}$ and the
associated gain in enthalpy decreases linearly with increasing
pressure, being maximum at the lowest accessible $P$ for the liquid,
and being zero at $P=J/v_{HB}$. For $P>J/v_{HB}$ the formation of hydrogen
bonds increases the total enthalpy. 

In the following we choose $J/\epsilon=0.5$, $J_\sigma/\epsilon
=0.05$, $v_{HB}/v_0=0.5$ as in Ref. \cite{FMS2003}, for sake of
comparison. In the next sections we describe the simulation method
adopted here and the results.

\section{Kawasaki Monte Carlo Method}

Monte Carlo (MC) simulations are performed in the 
ensemble with 
constant $N$, $P$, $T$ for  $N=1875$ water molecules distributed in
a mesh of $L^2=2500$ cells, corresponding to 75\% water
concentration. We do not observe appreciable size effects at constant
water concentration for  
($N=300$, $L^2=400$) and ($N=1200$, $L^2=1600$). Preliminary results
at 90\% water concentration display no significant differences with
the present case. 

Since we are interested in studying the dynamic behavior of the system when
diffusion is allowed, we adopt a MC method {\it a la}
Kawasaki \cite{Kawasaki}. This method consists in exchanging the
content of two stochastically selected nearest neighbors (n. n.) cells
and accepting the exchange with probability
\begin{equation}
\left\{
\begin{array}{ll}
\exp[-\Delta H/k_BT] & {\rm if} \ \Delta H>0 \cr
1		& {\rm if} \  \Delta H<0 \cr 
\end{array}\right.
\end{equation}
where $\Delta H $ is the difference between the enthalpy of the final
state and that of the initial state.

The same probability is adopted for every attempt to change the state
of any of the bond indices $\sigma_{ij}$.
One MC step consists of $5N$ trials of exchanging n. n. cells $i$
and $j$ or changing
the state of the bond indices $\sigma_{ij}$ (every time selecting at random which kind of
move and which $i$ and $j$ are considered) 
 followed by a volume
change attempt. In the latter case, 
the new volume is selected at random in the interval $[V-\delta V,
V+\delta V]$ with $\delta V=0.5v_0$.

For sake of simplicity, in the Lennard-Jones energy calculation
we do not consider the volume changes due to
the hydrogen bonds formation 
and we allow the formation of hydrogen bonds independently from the
size of the n. n. cells.
We include a cut-off for the Lennard-Jones interaction at a distance
equal to three times the size of a cell. Despite with this choice the cut-off is
density dependent, we tested that in this way we largely reduce
the computational cost  and  that the 
results do not change in a significant way with respect to the
 case without the cut-off.

In the simulations we follow an annealing protocol, by preparing a
random configuration and equilibrating it at high $T$ and low $P$ (gas
phase). We equilibrate the system over  $5\times 10^5$ MC steps, and
calculate the observable averaging over at least $2\times 10^6$ MC
steps. We use the final equilibrium configuration at a given
temperature as the starting configuration for the next lower temperature.

\section{Results and Discussion}

\subsection{The Phase Diagram}

\begin{figure}
\includegraphics[width=15cm]{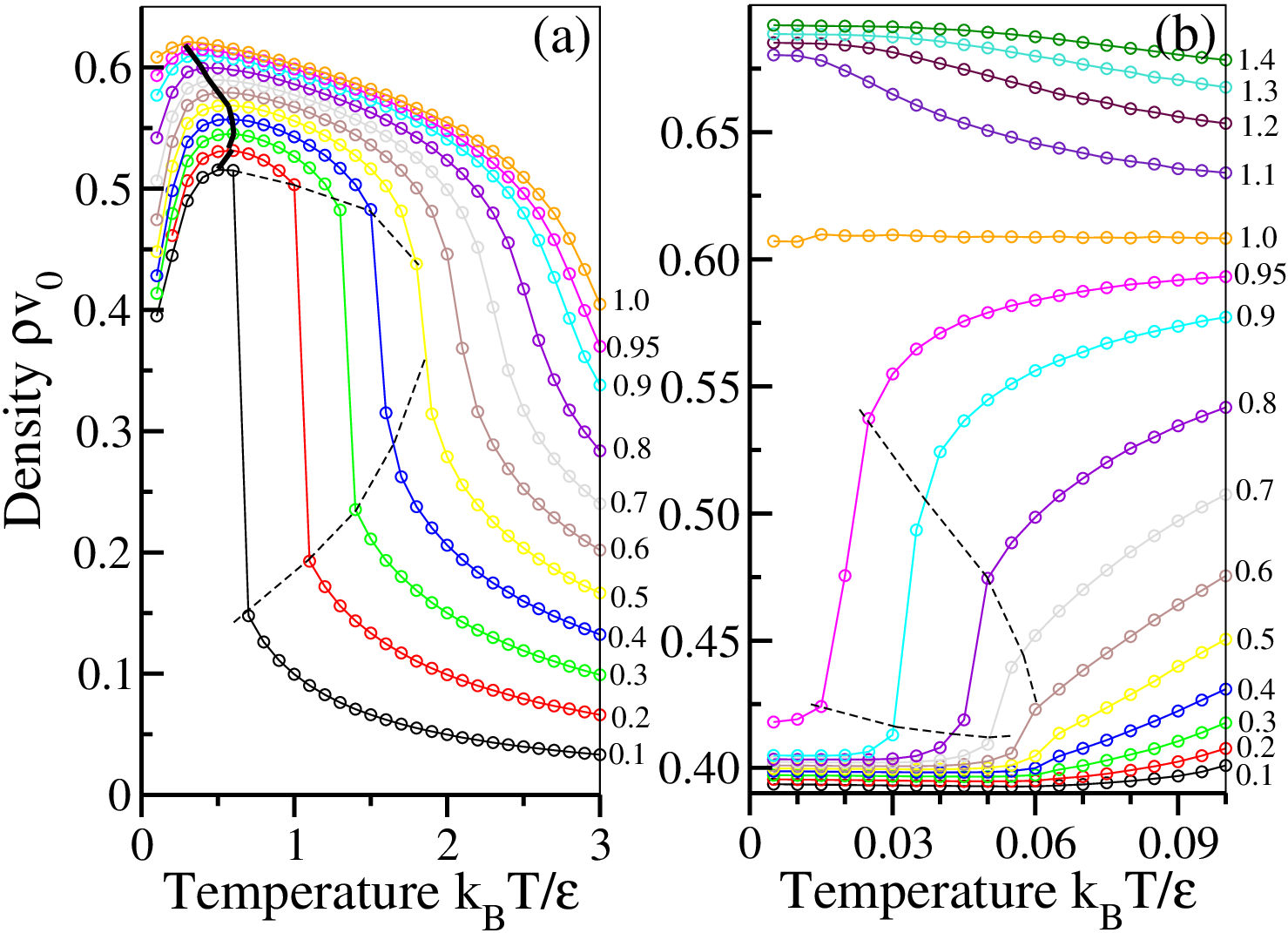}
\caption{Density $\rho$ of fluid water as function of temperature $T$ at
different values of pressure $P$. 
(a) At high $T$ a discontinuity in
$\rho$ appears below $Pv_0/\epsilon=0.5$, marking the first order phase transition
between gas (at low $\rho$) and liquid (at high $\rho$), ending in
a critical point at approximately $k_BT_C/\epsilon=1.9\pm 0.1$,
$P_Cv_0/\epsilon=0.50\pm 0.05$. In the liquid phase, isobars show a
temperature of 
maximum density (TMD line, marked by the thick black line) one of the
characteristic features of water. 
(b) At low $T$ the density rapidly decreases for
$P<J/v_{HB}=1\epsilon/v_0$ with a discontinuity marking the first
order phase transition
between high density liquid (at high  $\rho$) and low density liquid (at low $\rho$), ending in
a critical point at approximately $k_BT_{C'}/\epsilon=0.06\pm 0.01$,
$P_{C'}v_0/\epsilon=0.60\pm 0.15$. 
In both panels, symbols are results of the
MC calculation, lines are guides for the eyes, errors are comparable
to the size of the symbols, labels on the right of each panel are the
values of $Pv_0/\epsilon$ of the nearest curve. }
\label{fig1}
\end{figure}

From the calculation of the density of the system at fixed $P$ as
function of $T$ we find a liquid-gas first order phase transition
ending in a  critical point $C$ at approximately $k_BT_C/\epsilon=1.9\pm 0.1$,
$P_Cv_0/\epsilon=0.50\pm 0.05$ (Fig. \ref{fig1}a). In the liquid
phase, we observe a temperature of maximum 
density (TMD) for any isobar with
$Pv_0/\epsilon\leq 1$. This anomalous behavior in density is one of
the characteristic features of water. Here we note that $Pv_0/\epsilon\leq 1$
corresponds to the limiting value
$P=J/v_{HB}$ above which the formation of bonds increases the 
enthalpy in Eq.(\ref{p_term}), consistent with the intuitive idea that the density maximum
in water is a consequence of the open structure due to the hydrogen
bond formation. For $P>J/v_{HB}$ the formation of hydrogen bonds is no
longer convenient in terms of free energy and the system recovers the
behavior of a normal liquid (with no open structures). 

At lower $T$ and for $P\leq J/v_{HB}=1 \epsilon/v_0$, the isobars
display another discontinuity between high density and low density
(Fig.\ref{fig1}b), both at $\rho$  higher than the gas density. As in the case with no
diffusion  \cite{FMS2003}, we associate this
discontinuity in $\rho$ to a first order phase transition between a high density
liquid (HDL) and a low density liquid (LDL) 
ending in critical point $C'$ at $k_BT_{C'}/\epsilon=0.06\pm 0.01$ and
$P_{C'}v_0/\epsilon=0.60\pm 0.15$. 
The values of the critical temperature and pressure for $C$ and the
critical temperature for $C'$ are consistent, within the error bars, to those previously estimated
\cite{FMS2003,FS2007}, while the value of $P_{c'}$ is somehow
smaller. This is due to the difficulty in estimating the position of
$C'$ from the discontinuity of $\rho$ and it will be discussed in a
future work. Here, instead, we focus on the analysis of the dynamics of the
system at low $T$.  

\subsection{The Dynamics at Low Temperature}

To get an insight into the processes at low $T$ associated
with the  hydrogen bond dynamics, we calculate the 
the quantity $M_i\equiv\frac{1}{4}\sum_j\sigma_{ij}$,
that gives the orientational order of the four bond indices of
molecule $i$, and its correlation function 
\begin{equation}
C_M(t)\equiv\frac{1}{N}\sum_i
\frac{\langle M_i(t_0+t)M_i(t_0)\rangle-\langle M_i \rangle^2}
{\langle M_i^2 \rangle-\langle M_i \rangle^2} 
\end{equation}
where the $\langle \cdot \rangle$ stands for the thermodynamic
average \cite{KFS2008,MSSSF2009}.

As we have discussed in the previous section, the effect of the
hydrogen bond is evident only at pressure $P\leq J/v_{HB}$ and below
the TMD line.
Therefore we focus on state points at $P\leq 1 \epsilon/v_0$ and
$T\leq 0.4\epsilon/k_B$.

\begin{figure}
\includegraphics[width=15cm]{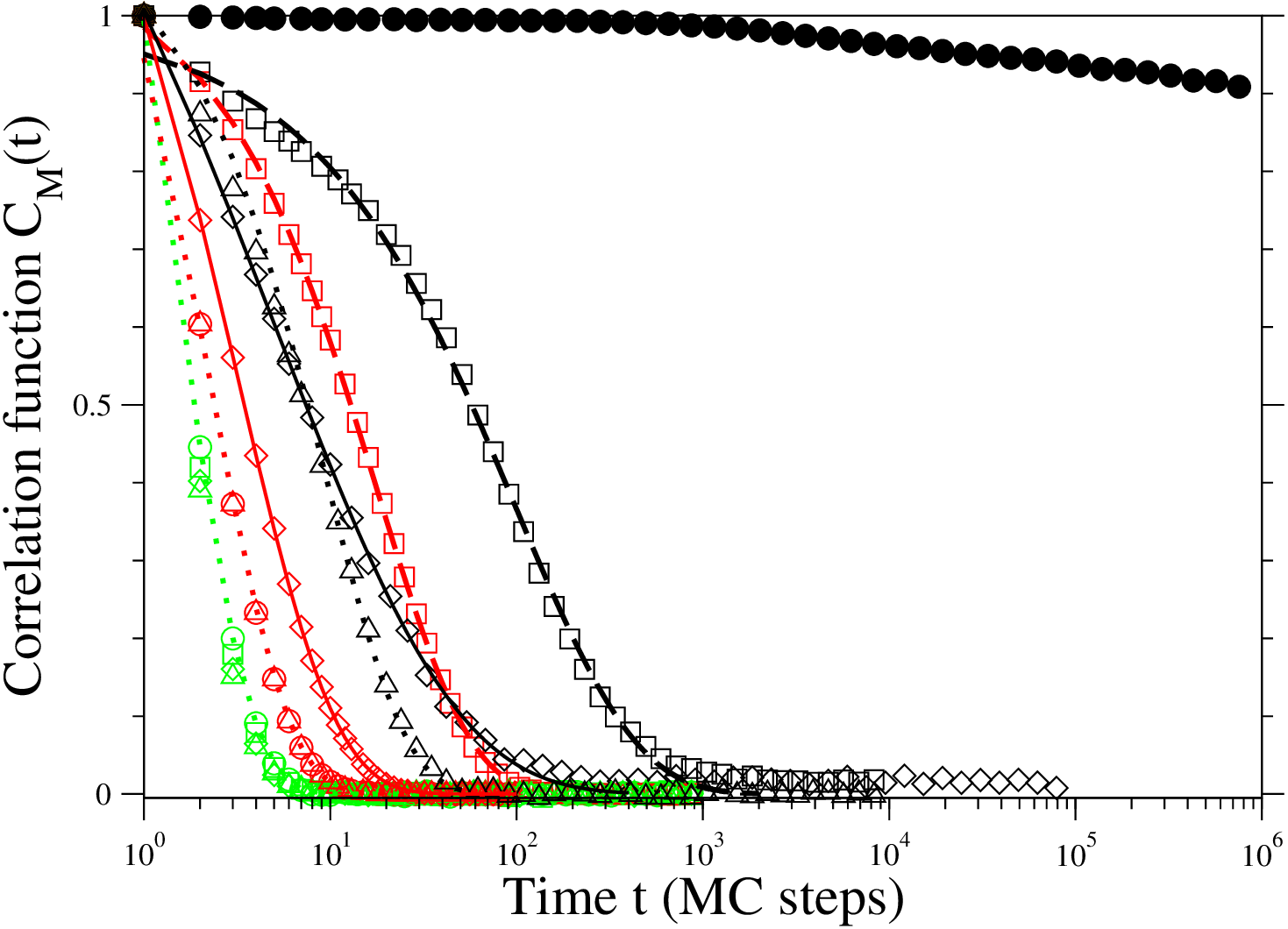}
\caption{The correlation function $C_M(t)$ for different pressure and
  temperature below the temperature of maximum density (TMD). The
  correlation decays as an exponential (dotted lines) at any $T\geq 0.4\epsilon/k_B$,
just below the TMD, for any $P<1.0 \epsilon/v_0$ (green symbols)
and for all $T$ at $P=1\epsilon/v_0$ (Triangles). 
The correlation $C_M(t)$ 
is well described by a stretched exponential (see text) for $P=0.7
\epsilon/v_0$ (Diamonds) and $P=0.4
\epsilon/v_0$ (Squares) for 
$T\leq 0.1\epsilon/k_B$ with a stretched exponent $\beta$ that is larger at
lower $T$. At low pressure $P=0.1 \epsilon/v_0$ (Circles), $C_M(t)$  is
non-exponential only at very low $T=0.05\epsilon/k_B$ (full Circles).} 
\label{fig2}
\end{figure}

\begin{figure}
\includegraphics[width=5cm]{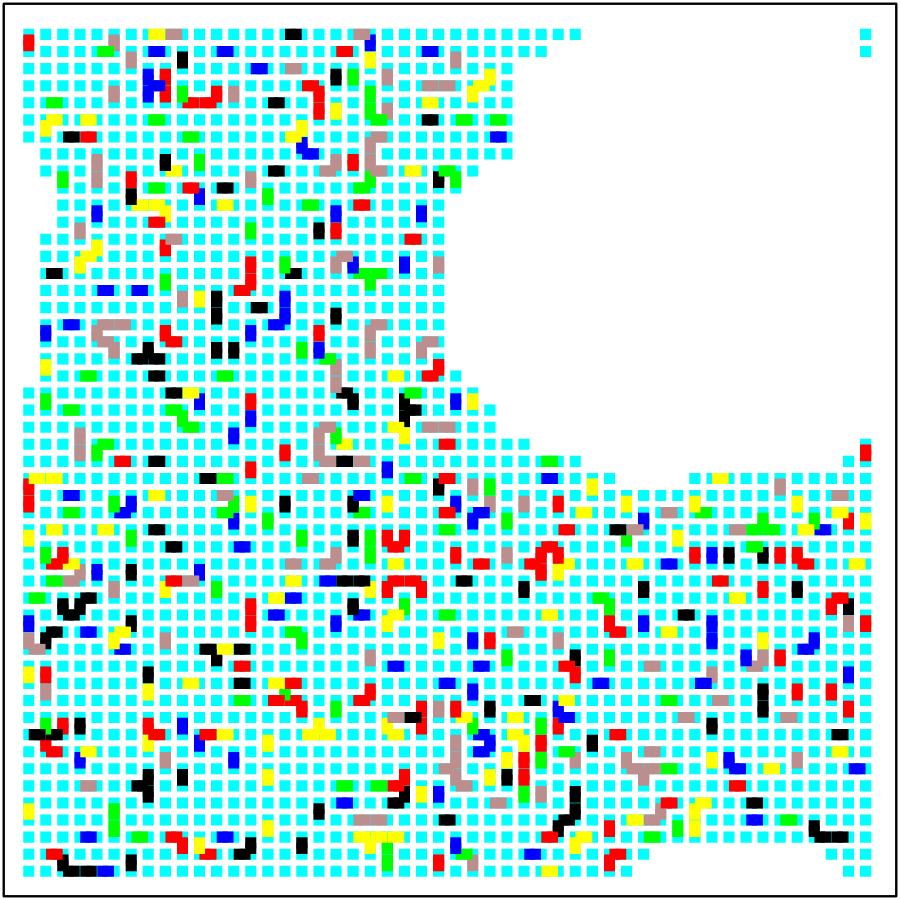}
\includegraphics[width=5cm]{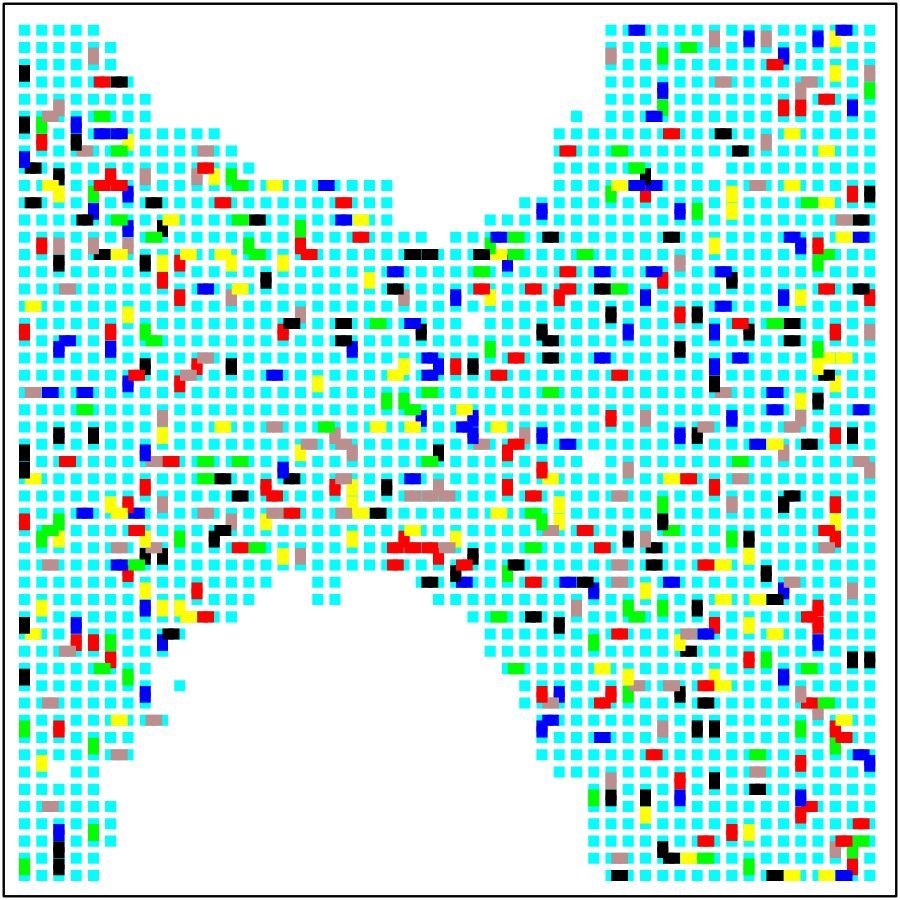}
\includegraphics[width=5cm]{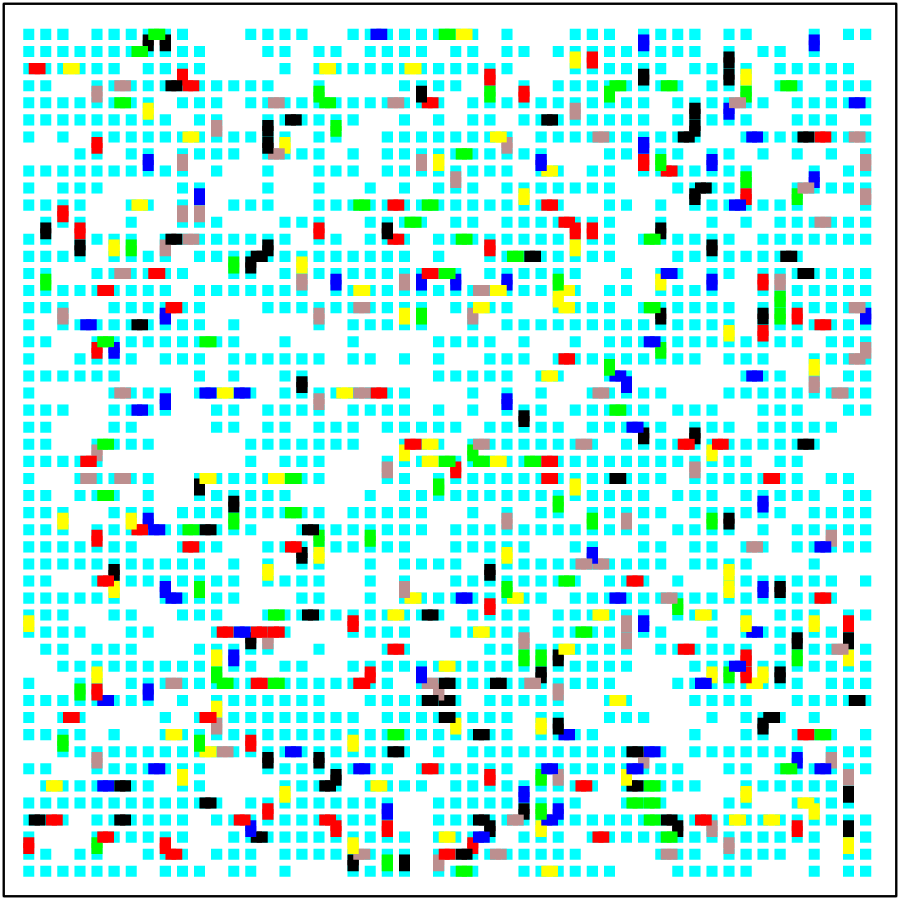}
\includegraphics[width=5cm]{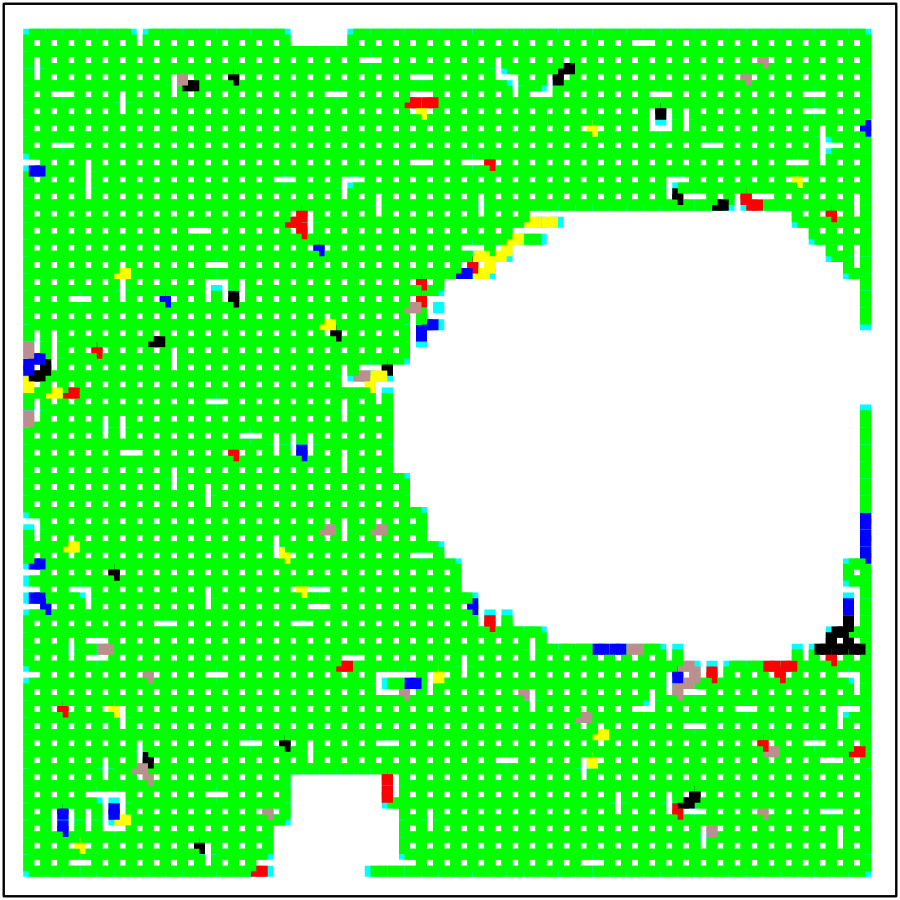}
\includegraphics[width=5cm]{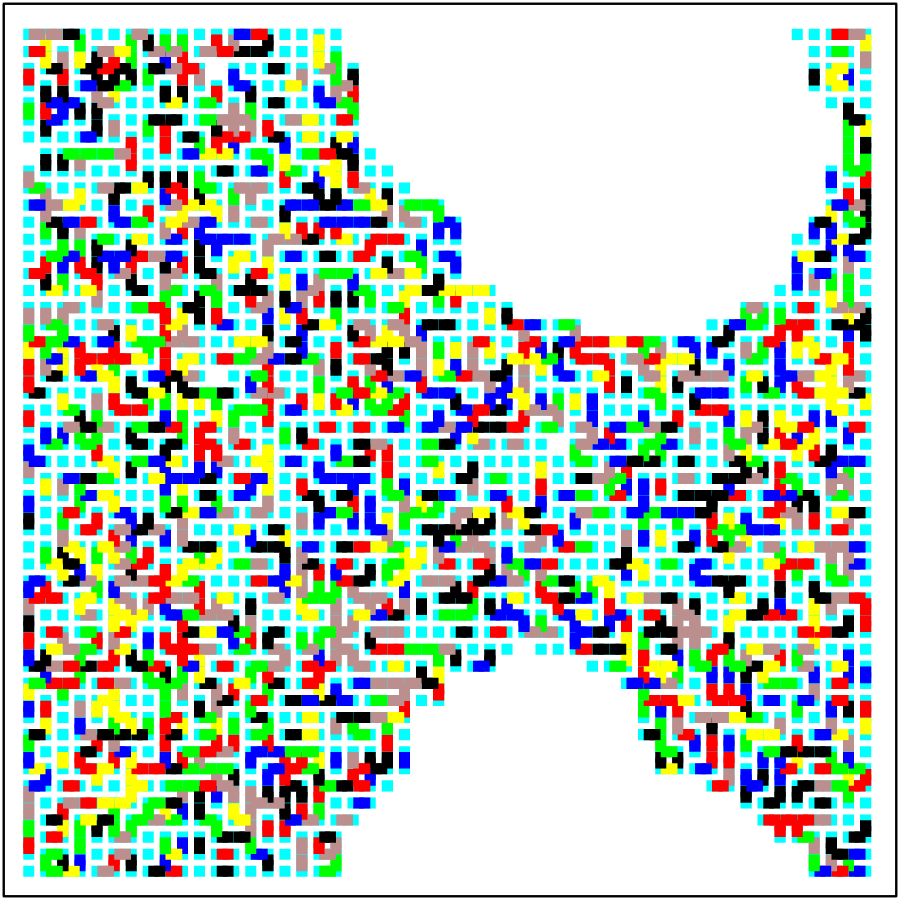}
\includegraphics[width=5cm]{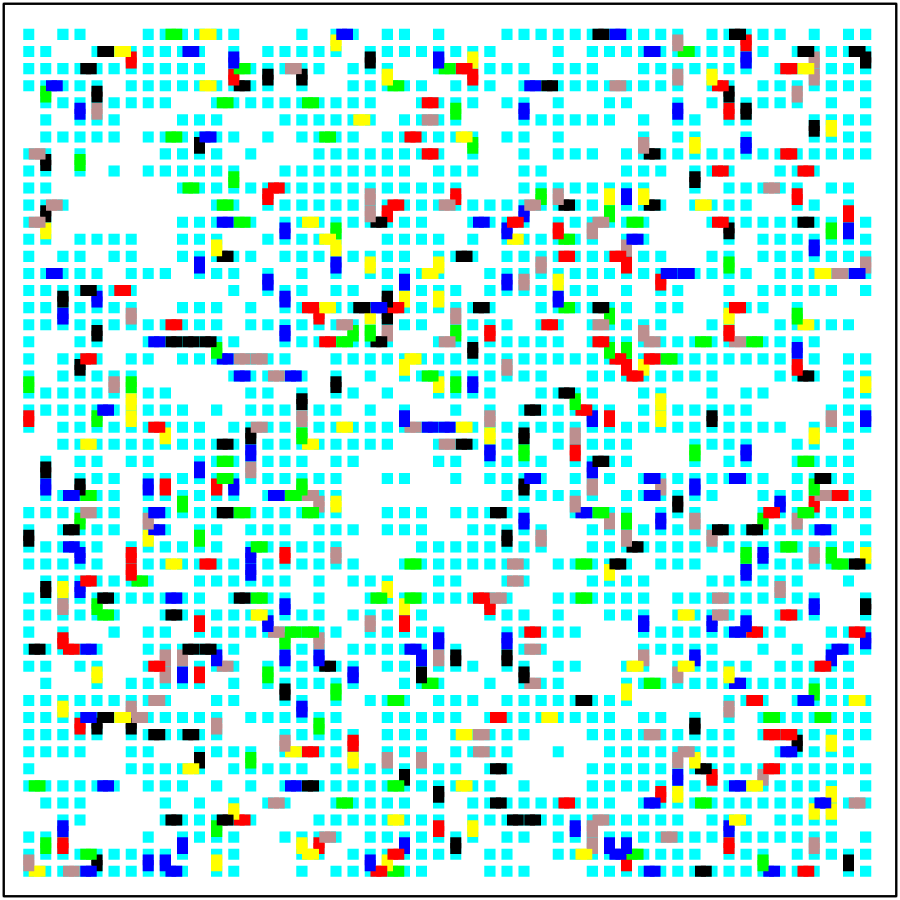}
\includegraphics[width=5cm]{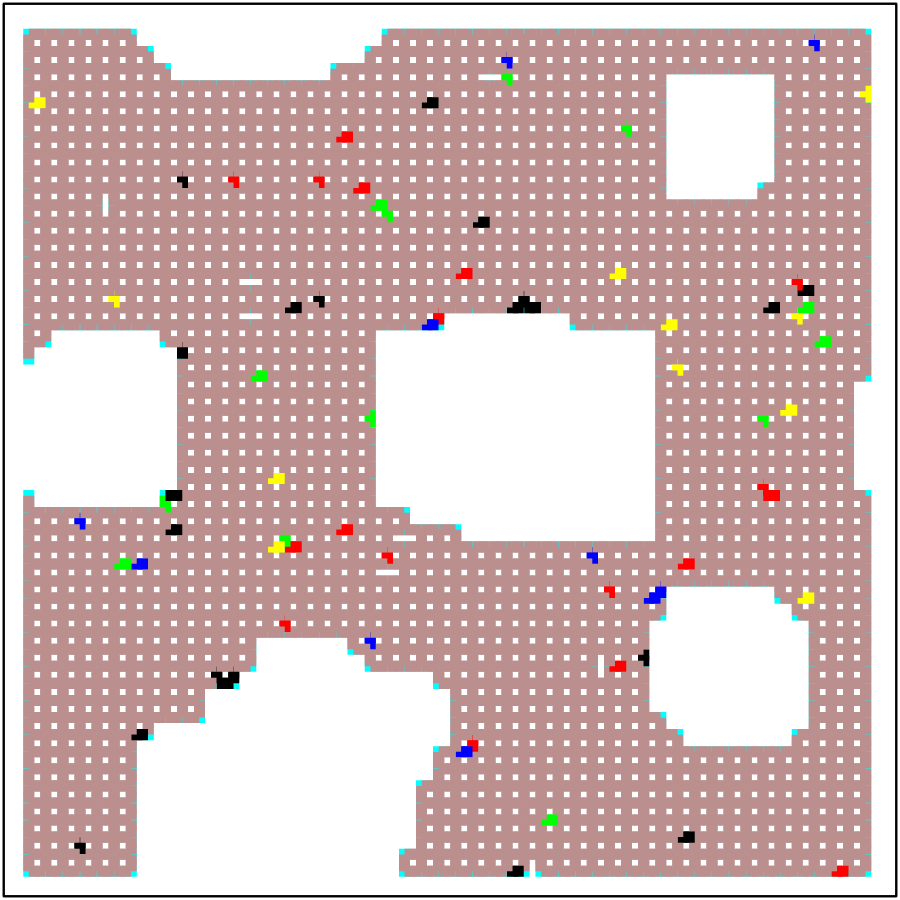}
\includegraphics[width=5cm]{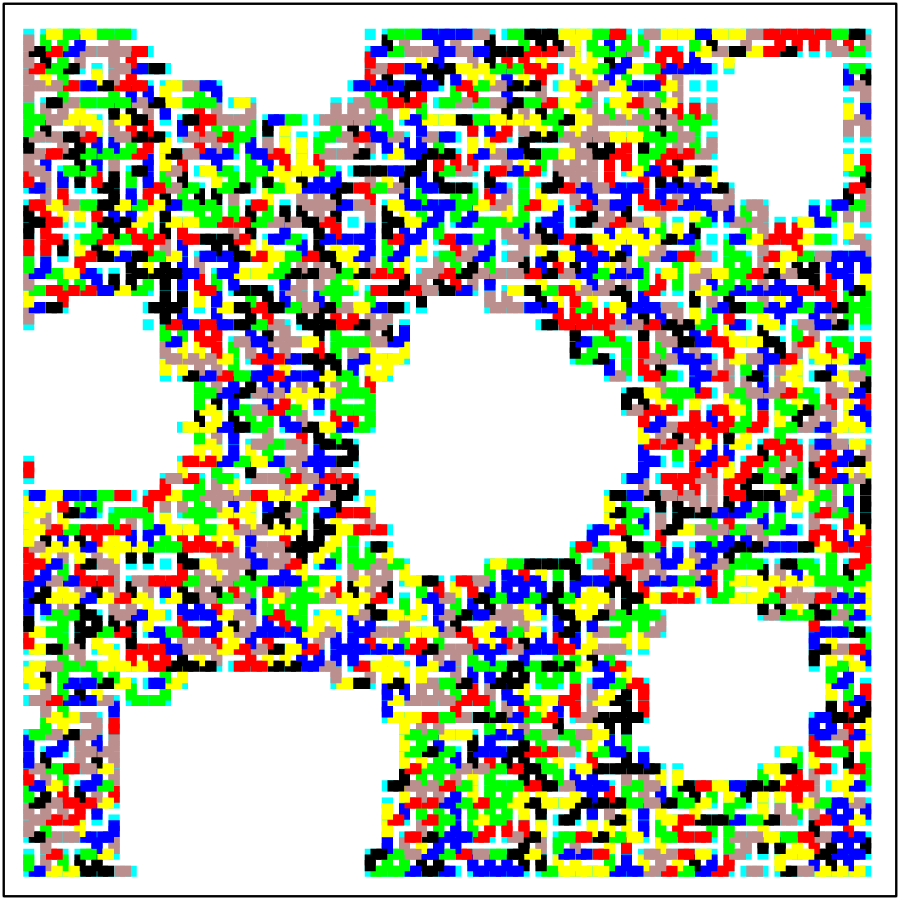}
\includegraphics[width=5cm]{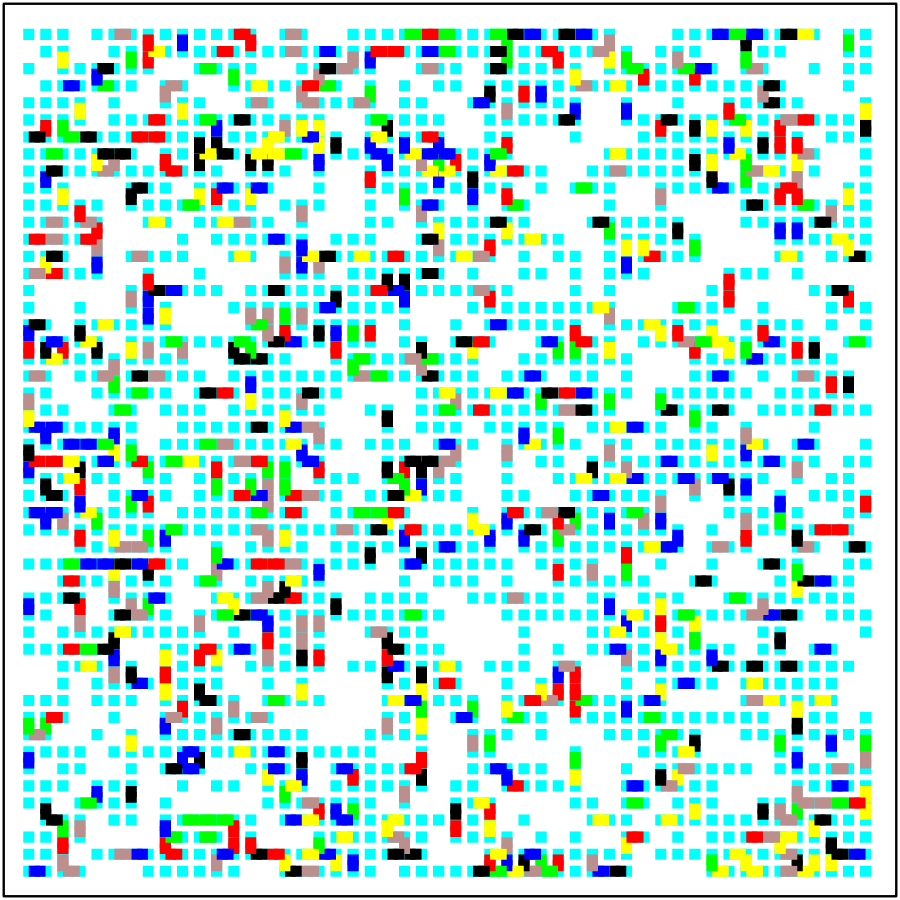}
\includegraphics[width=5cm]{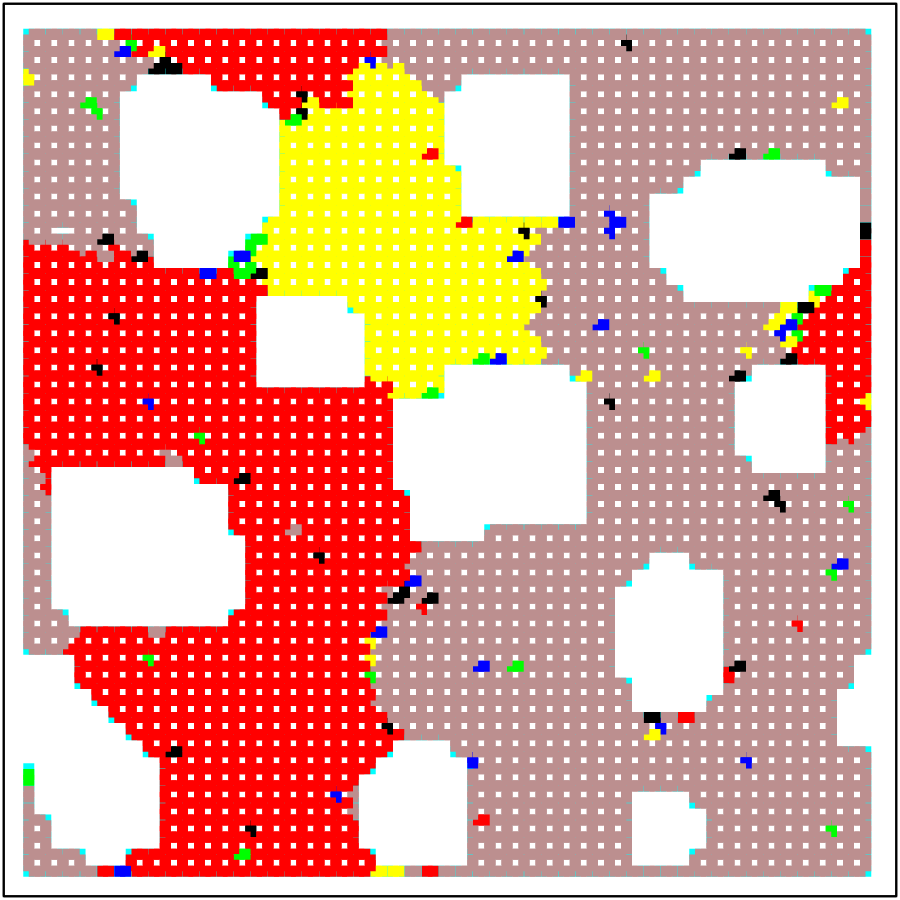}~~
\includegraphics[width=5cm]{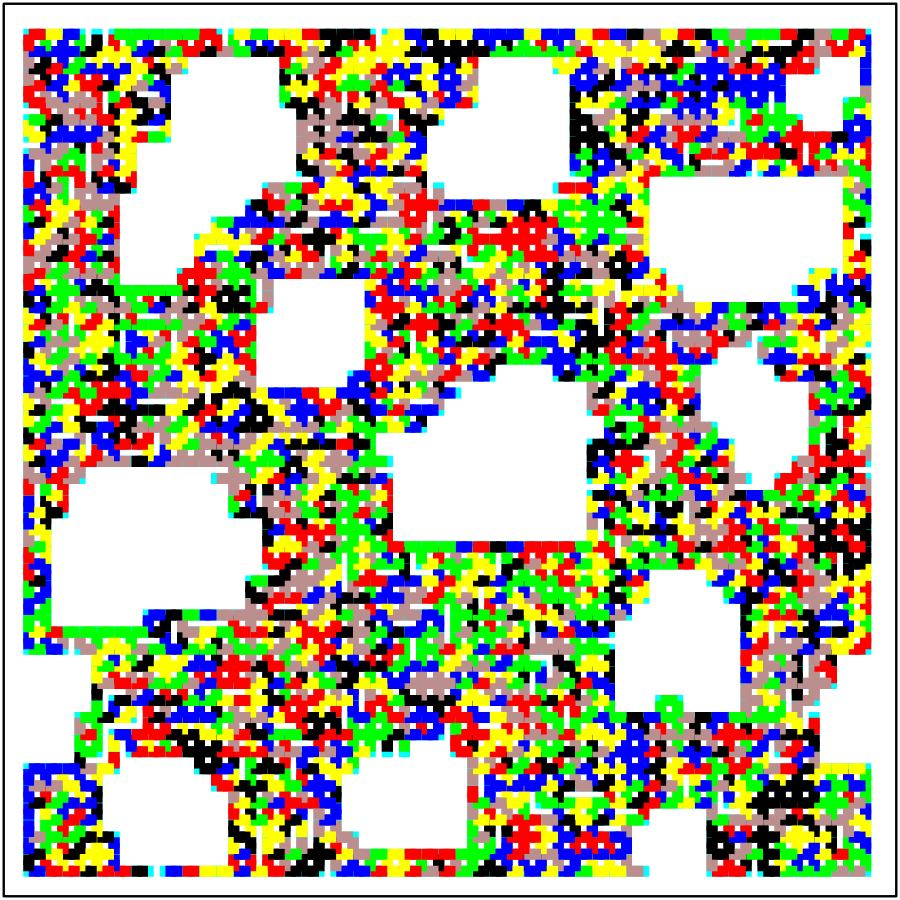}~~
\includegraphics[width=5cm]{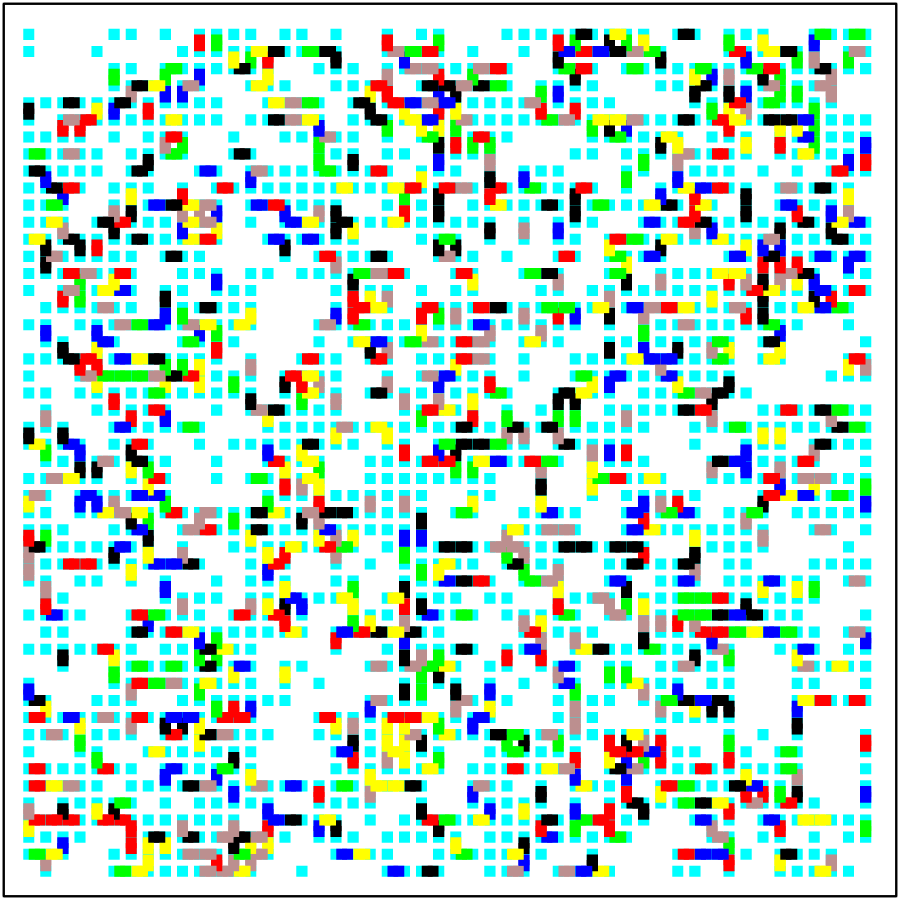}
\caption{Hydrogen bonds configurations 
  showing the hydrogen bond ordering at low $T$ and at $P<
  J/v_{HB}=1\epsilon/v_{HB}$. 
Symbols: Turquoise dots represent cells containing water molecules
with at least one bond free;
colored lines represent hydrogen bonds between molecules in n. n. cells, with
six possible colors corresponding to the $q=6$ possible values of the
orientational state.
Every row corresponds to a different $P$: from top,  
$P=1 \epsilon/v_0$,
$P=0.7 \epsilon/v_0$,
$P=0.4\epsilon/v_0$,
$P=0.1 \epsilon/v_0$.
Every column corresponds to a different $T$: from  right,
$T=0.4\epsilon/k_B$,
$T=0.1\epsilon/k_B$, 
$T=0.05\epsilon/k_B$.
The size of the box is not in scale with the total volume of the
system at the different values of $P$ and $T$.}
\label{fig3}
\end{figure}

\begin{itemize}

\item
We first observe that for $P=1 \epsilon/v_0$ the correlation
function is always  exponential for all the temperatures with
$0.005 \leq k_BT/\epsilon \leq 0.4$ (Triangles in Fig.\ref{fig2}).
From the exponential fits (dotted lines in Fig.\ref{fig2})
the characteristic decay time is 
$\tau\simeq 1.3$ MC Steps (MCS) at $k_BT/\epsilon =0.4$ (green symbols), 
$\tau\simeq 2.2$ MCS at $ k_BT/\epsilon =0.1$ (red symbols), 
$\tau\simeq 10$ MCS at $ k_BT/\epsilon =0.05$ (black symbols). 
 Therefore, the correlation time increases of one order of magnitude
 for a decrease of $T$ of one order of magnitude,  but $\tau$ is
 always much shorter than the simulation time.

This behavior is consistent with the observation of the typical
configurations of hydrogen bonds at these $P$ and $T$ (upper row in
Fig.~\ref{fig3}). The water molecules condense on the hydrophobic
surface, leaving a large dehydrated area, but the number of hydrogen
bonds does not increase in an evident way and no orientational order is
observed even at the lowest $T$.

\item
Next we decrease the pressure to $P=0.7 \epsilon/v_0$ (Diamonds in
Fig.\ref{fig2}), close to the the liquid-liquid critical pressure. 
At $k_BT/\epsilon =0.4$ (green symbols), 
the correlation function is exponential with $\tau\simeq 1$ MCS. 
At lower $T$ 
the correlation function is no longer exponential.
It can be well described with a stretched exponential function
\begin{equation}
C_M(t)=C_0 \exp\left[-\left(t/\tau\right)^\beta \right]
\end{equation}
where $C_0$, $\tau$ and $\beta\leq 1$ are fitting constant. 
For $\beta=1$ the function is exponential, and the more stretched is
the function, the smaller is the power
$0<\beta\leq 1$.

At $k_BT/\epsilon =0.1$  (red symbols), 
the stretched exponential (continuous red line) 
has $\beta=0.8$ and $\tau\simeq 3.1$ MCS, while at 
$k_BT/\epsilon =0.05$  (black symbols), 
the stretched exponential (continuous black line) 
has $\beta=0.4$ and $\tau\simeq 5$ MCS.

Hence, along isobars close to the liquid-liquid critical pressure, the
dynamics of the liquid is not much slower than that at
$Pv_0/\epsilon=1$, but it becomes non exponential below the TMD, a
typical precursor phenomenon of glassy systems  \cite{FC1999}.
This phenomenology is usually associated with the presence of
heterogeneity in the system. While in a homogeneous system the
characteristic relaxation time is determined by some characteristic
energy scale, in a heterogeneous system the presence of many
characteristic energies induces the occurrence of many typical time
scales, giving rise to a non exponential behavior.

This is consistent with the increasing number of hydrogen bonds that
can be observed at this pressure in the high density liquid phase 
(central panel on the second row in Fig.~\ref{fig3}). The formation of
many hydrogen bonds inhibits the free diffusion of the molecules,
determining heterogeneities.

This effect is extremely evident at the lower $k_BT/\epsilon =0.05$,
where the non exponential behavior of the correlation function is very
strong ($\beta=0.4$) and the hydrogen bond network is fully developed,
but with many isolated locally disordered regions
(left panel on the second row in Fig.~\ref{fig3}). This is the typical
configuration of the hydrophobic surface at low $T$  at this $P$. The
surface is partially hydrated by low density liquid water forming an 
open network of bonded molecules. The network is
almost complete, but many small heterogeneities are present.

\item
By further decreasing the pressure to $P=0.4\epsilon/v_0$ (Squares in
Fig.\ref{fig2}), we observe an exponential decay of the correlation
function at $k_BT/\epsilon =0.4$ (dotted line) with $\tau\simeq 1.3$ MCS
and a stretched exponential behavior when the liquid-liquid critical
temperature is approached. Differently from the case at
$P=0.7\epsilon/v_0$, at $P=0.4\epsilon/v_0$ the stretched exponent
does not decrease in a strong way, but the relaxation time
largely increases at lower $T$.

In particular, 
at $k_BT/\epsilon =0.1$  (red symbols), 
the stretched exponential (dashed red line) 
has $\beta=0.9$ and $\tau\simeq 18$ MCS, while at 
$k_BT/\epsilon =0.05$  (black symbols), 
the stretched exponential (dashed black line) 
has $\beta=0.7$ and $\tau\simeq 100$ MCS. Therefore, the correlation time
slows down of two order of magnitudes by decreasing the $T$ from 
$k_BT/\epsilon =0.4$ to $k_BT/\epsilon =0.05$.

This different behavior with respect to the case at
$P=0.7\epsilon/v_0$ could be associated to the different evolution of
the hydrogen bond network. Previous mean field \cite{FS2007} and MC
results \cite{KumarFS2008} have shown that the increase of the number
of hydrogen bonds with decreasing $T$ is much more gradual at low $P$
than at high $P$. Therefore, at high $P$ this process could
generate quenched-in heterogeneities and strong non-exponentiality. 
At low $P$, instead, the gradual building up of the hydrogen bond
network freezes the system, inducing a large slowing down. As a
consequence, the systems has large relaxation times, but the hydrogen
bond network develops in a more ordered way 
(third row in Fig.~\ref{fig3}). 

The gradual increase of the number of hydrogen bonds is also responsible
for the slower dehydration of the hydrophobic surface. Many
small dehydrated regions  seem to be annealed at a very slow rate
by the progressive network formation. 

\item
We finally consider the lowest pressure $P=0.1\epsilon/v_0$ (Circles in
Fig.\ref{fig2}).
In this case, the correlation time is exponential for  $k_BT/\epsilon
=0.4$ (dotted green line) and  $k_BT/\epsilon =0.1$ (dotted red line),
with $\tau\simeq 1.3$ MCS and $\tau\simeq 2.2$ MCS, respectively.

At these temperatures the system is homogeneous, with
no order at $k_BT/\epsilon=0.4$ (right panel on the bottom  row in Fig.~\ref{fig3})
and with small orientationally ordered regions at  $k_BT/\epsilon =0.1$ 
(central panel on the bottom  row in Fig.~\ref{fig3}). These regions
are the effect of the slowly increasing number of hydrogen bonds
$N_{HB}$ and are possible only at low $P$, because only at low values
of $P$  the enthalpy increase proportional to $N_{HB}$ [Eq.(\ref{p_term})]
is small.

As a consequence of the gradual increase of these locally ordered
regions of bonded molecules, at lower $T$ the system is trapped in a
metastable excited state where large regions of ordered bonded
molecules, with different orientations, compete in their growth
process (left panel on the bottom  row in Fig.~\ref{fig3}). These
process resembles the arrested dynamics of other model systems for
glassy dynamics \cite{Dawson2002}.
It is characterized by non exponential correlation behavior with
characteristic times that are several order of magnitudes larger than
those at slightly higher $T$. In our case the correlation time largely
exceeds the simulation time of $10^6$ MCS (Full Circles  in Fig.~\ref{fig2}).

\end{itemize}

\section{Conclusions}

We have studied the dynamics of the hydrogen bond network of
a monolayer of water between hydrophobic plates at partial hydration. 
We adopt  a model for water with cooperative interactions 
\cite{FS-PhysA2002,FS2002,dlSF2009}.
The model can reproduce the different scenarios proposed for water
\cite{Stokely08}. We consider the parameters that give rise to a
liquid-liquid phase transition ending in a critical point
\cite{FMS2003,FS2007}. It has been shown that in the supercooled
regime, below the TMD line, the
system has slow dynamics \cite{KFS2008,KumarFS2008,MSSSF2009}.
Here we adopt a diffusive MC dynamics to study the case at partial
hydration. 

Our results show a complex phenomenology. At high $P$ the effect of
the hydrogen bond network is negligible due to the high increase of
enthalpy for the hydrogen bond formation. This is reflected in the
dynamics by a moderate increase of the  correlation time also at very
low $T$. Therefore, the system reaches the equilibrium easily.
 As a consequence, we typically observe the formation of a
single large dehydrated region. 

At a pressure close to the liquid-liquid critical pressure, the rapid 
formation of the hydrogen bond network determines the presence of
large amount of heterogeneity in the system. As a consequence the
correlation function shows a strong non-exponential behavior at low
$T$, where we typically observe a well developed hydrogen bond
network with many local heterogeneities. Nevertheless, the hydrogen
bond network is fully developed only at very low $T$, keeping the
correlation time short and allowing the formation of large dehydrated
regions on the hydrophobic surface.

At lower $P$, the number of hydrogen bonds grows in a gradual and more
homogeneous way when the temperature decreases. This process slows
down the correlation decay of orders of magnitudes. Nevertheless,  
the effect on the exponential character of the correlation function is
small, because the system has only a small amount of heterogeneity. 
However, the effect on the process of dehydration of the surface is 
evident.

This effect is extremely strong at very low $P$, where 
regions of ordered bonded molecules with different orientation grow
in a competing fashion. This process give rise to an arrested dynamics,
typical of many glassy systems, with the correlation time that
increases of many orders of magnitude for a small decrease in
temperature. The dehydration process in this case is almost completely frozen on the
time scale of our simulations. 

\section*{Acknowledgments}
We thank our collaborators, 
P. Kumar,
M. I. Marqu\'es,
M. Mazza, 
H. E. Stanley,
K. Stokely,
E. Strekalova, 
for discussions on different aspects of this topic.
This work was partially supported by the Spanish Ministerio de Cienci
a
y Tecnolog\'ia 
Grants Nos. FIS2005-00791 and FIS2007-61433, Junta de Andaluc\'ia
Contract No. FQM 357.

\section*{References}

\bibliographystyle{./iopart-num}

\end{document}